\documentclass[12pt]{iopart}

\usepackage{iopams} 
\expandafter\let\csname equation*\endcsname\relax
\expandafter\let\csname endequation*\endcsname\relax
\usepackage{amssymb,amsmath}

\usepackage{graphicx}
\usepackage{hyperref}
\usepackage{cite} 

\begin{document}

\title[Second quantization approach to COVID-19 epidemic]{Second quantization approach to COVID-19 epidemic}

\author{Leonardo Mondaini$^1$, Bernhard Meirose$^2$ and Felipe Mondaini$^3$}
\address{$^1$ Department of Physics, Federal University of the State of Rio de Janeiro, Rio de Janeiro, Brazil}
\address{$^2$ Fysiska institutionen, Lunds universitet, Lund, Sweden}
\address{$^3$ Centro Federal de Educação Tecnológica Celso Suckow da Fonseca, CEFET-RJ/ Campus Petrópolis, 25.620-003, Petrópolis, Brazil}

{\small 
\begin{center}
*All authors contributed equally to this work.
\end{center}
}

\eads{\mailto{mondaini@unirio.br},\mailto{bernhard.meirose@hep.lu.se},\mailto{felipe.mondaini@cefet-rj.br}}
\vspace{10pt}
\begin{indented}
\item[]December 2021
\end{indented}

\begin{abstract}
In this article, a stochastic SIR-type model for COVID-19 epidemic is built using the standard field theoretical language based on creation and annihilation operators. From the model, we derive the time evolution of the mean number of infectious (active cases) and deceased individuals. In order to capture the effects of lockdown and social distancing, we use a time-dependent infection rate. The results are in good agreement with the data for three different waves of epidemic activity in South Korea.

\end{abstract}

%
\vspace{2pc}

\section{Introduction}

The year 2020 will be remembered in history as the year in which humanity suffered from the rapid spread of coronavirus disease 2019 (COVID-19), caused by the severe acute respiratory syndrome coronavirus 2 (SARS-CoV-2) and responsible for the worst pandemic in the last hundred years. Governments around the world were strongly impacted by the pandemic and, in addition to efforts of a social and economic nature, it became evident the need to bring together members of the international scientific community to analyze data on the evolution of the disease and propose concrete actions for the control and mitigation of its serious effects~\cite{law2020tracking}. In this sense, even researchers from areas far from epidemiology were encouraged to collaborate with their experience and specific knowledge in the study of possible models for describing the evolution of the pandemic in different countries of the world~\cite{thomas2020primer,zhuang2020preliminary,zhao2020preliminary,neves2020predicting,medo2020contact}, often serving as guides for the actions of local governments. Theoretical particle physicists are used to dealing with problems and proposing models to describe complex phenomena involving the fundamental constituents of matter. In this work, we show how the standard field theoretical language based on creation and annihilation operators (building blocks of the second quantization method \cite{Greiner}), usually used in the construction of these elementary particle models, may also be adapted to the construction of an epidemiological model for COVID-19 which can be compared to the data available for several countries\footnote{M. Doi\cite{doi1976} was the first to point that, despite what one could infer from the name, the second \emph{quantization} approach is not limited to \emph{quantum} systems. Indeed, this approach may be introduced to the description of certain \emph{classical} many-particle systems, such as the one we are dealing with in this work.}. We believe this can serve as an stimulus so that other researchers working on these subjects can also contribute to this great scientific network. Indeed, in a recent letter \cite{MondainiBMST2015}, we have already shown how this field theoretical language may be used to derive closed master equations \cite{Gardiner} describing the population dynamics of multivariate stochastic epidemic models. Since then we applied a simplified version of the theoretical framework established in Ref.~\cite{MondainiBMST2015} to the description of tumour growth \cite{Mondaini_2017}. Nevertheless, no successful confrontation with real data was possible on both works.

In what follows we adapt the methodology established in Ref.~\cite{MondainiBMST2015} so as to derive an SIR-type stochastic model for COVID-19 epidemic, from which we obtain the time evolution of the mean number of infectious (active cases) and deceased individuals in South Korea. Our choice for South Korea was motivated by its mass testing and careful monitoring of infected patients since appearance of their first symptoms \cite{dongarwar2021implementation}. This leads us to the first successful description of real data and allows us to finally confirm the robustness and versatility of the approach. Indeed, our main motivation comes from the fact that, as remarked in Ref.~\cite{dodd2009many}, for the types of models studied in population biology and epidemiology, this field theoretical description is notationally neater and more manageable than standard methods, often replacing sets of equations by single equations with the same content. As an example of this, we may stress that a single Hamiltonian function sums up the system dynamics compactly and may be easily written down from a verbal description of the transitions presented in our model. 
Besides the simplicity and notational appeal, we may stress as another advantage of our SIR-type model, the use of a time-dependent infection rate in order to capture the effects of a lockdown, since the standard SIR model does not take these into account. This approach is similar to the one adopted in Ref.~\cite{palladino2020modelling}, although the infection rate function we use differs from Ref.~\cite{palladino2020modelling} at the initial stage of epidemic, i.e., before lockdown and social distance measures are applied.

The rest of this work is organized as follows. In Section \ref{model}, we introduce the basic aspects of our model which allow us to obtain differential equations describing the time evolution of the mean number of infectious and deceased individuals, while the time-dependent infection rate used for data comparison is presented in Section \ref{modified}. The differential equations are used to fit available data (South Korea) for the time evolution of infectious (active cases) individuals in Section \ref{active} and deceased individuals in Section \ref{dead}. Further data comparisons for two other distinct waves of epidemic activity in South Korea are presented in Section \ref{23outbreak}, whereas a discussion of the results is presented in Section \ref{discussion}. 
Our conclusions are presented in Section \ref{conclusion}.

\section{Building the model}
\label{model}

We start by considering interacting populations, whose total sizes are allowed to change, composed of four types of individuals: susceptible, infectious, recovered and deceased.
Let us introduce $\mathcal{S}(t)$, $\mathcal{I}(t)$, $\mathcal{R}(t)$ and $\mathcal{D}(t)$ as random variables which represent, respectively, the number of susceptible, infectious, recovered and deceased individuals at a given time instant $t$.

We then consider a multivariate process $\{(\mathcal{S}(t);\,\mathcal{I}(t);\,\mathcal{R}(t);\,\mathcal{D}(t))\}_{t=0}^\infty$ with a joint probability function given by

\begin{equation}
p_{(n_S,n_I,n_R,n_D)}(t)={\rm{Prob}}\{\mathcal{S}(t)=n_S;\,\mathcal{I}(t)=n_I;\,\mathcal{R}(t)=n_R;\,\mathcal{D}(t)=n_D\}~. \label{eq1}
\end{equation}

\vskip \baselineskip
Our aim is to compute time-dependent expectation values of the observables $\mathcal{S}(t)$, $\mathcal{I}(t)$, $\mathcal{R}(t)$ and $\mathcal{D}(t)$, which may be defined in terms of the configuration probability according to

\begin{eqnarray}
\left<\mathcal{S}(t)\right>=\sum_{n_S,n_I,n_R,n_D}n_S\,p_{(n_S,n_I,n_R,n_D)}(t)~;\nonumber\\
\left<\mathcal{I}(t)\right>=\sum_{n_S,n_I,n_R,n_D}n_I\,p_{(n_S,n_I,n_R,n_D)}(t)~;\nonumber\\
\left<\mathcal{R}(t)\right>=\sum_{n_S,n_I,n_R,n_D}n_R\,p_{(n_S,n_I,n_R,n_D)}(t)~;\nonumber\\
 \left<\mathcal{D}(t)\right>=\sum_{n_S,n_I,n_R,n_D}n_D\,p_{(n_S,n_I,n_R,n_D)}(t)~. \label{eq2}
\end{eqnarray}

\vskip \baselineskip
The probabilistic state of the system may be represented by the vector

\begin{equation}
\left|\nu_S,\nu_I,\nu_R,\nu_D\right>_t=\sum_{n_S,n_I,n_R,n_D} p_{(n_S,n_I,n_R,n_D)}(t)\left|n_S,n_I,n_R,n_D\right>~, \label{eq3}
\end{equation}

\vskip \baselineskip
\noindent
with the normalization condition $\sum_{n_S,n_I,n_R,n_D} p_{(n_S,n_I,n_R,n_D)}(t)=1$.


The configurations are fully given in terms of occupation numbers ($n_S,n_I,n_R,n_D$), thus a representation in terms of second-quantized bosonic operators \cite{Cardy_1998} seems natural. We introduce, therefore, {\it creation} and {\it annihilation} operators for the susceptible, infectious, recovered and deceased individuals, respectively, $a_S$ and $a^\dagger_S$, $a_I$ and $a^\dagger_I$, $a_R$ and $a^\dagger_R$, $a_D$ and $a^\dagger_D$. 

These operators must satisfy the following commutation relations

\begin{eqnarray}
\left[a_i,\,a^\dagger_j\right] = \delta_{ij}\,~;\nonumber \\
\left[a_i,\,a_j\right] = \left[a^\dagger_i,\,a^\dagger_j\right] = 0\,~, \label{eq4}
\end{eqnarray}

\vskip \baselineskip
\noindent
where $i,j=\{S,I,R,D\}$ and $\delta_{ij}$ is the Kronecker delta ($\delta_{ij}=1$ if $i=j$ and $\delta_{ij}=0$ if $i\neq j$). As usual in the second quantization framework, we say that $a^\dagger_S$, $a^\dagger_I$, $a^\dagger_R$ and $a^\dagger_D$ ``create" , respectively, susceptible, infectious, recovered and deceased individuals when applied over the reference (vacuum) state $\left|0,0,0,0\right>$. This allows us to build our space from basis vectors of the form $\left|n_S,n_I,n_R,n_D\right>=\left(a^\dagger_S\right)^{n_S}\left(a^\dagger_I\right)^{n_I}\left(a^\dagger_R\right)^{n_R}\left(a^\dagger_D\right)^{n_D}\left|0,0,0,0\right>$.

This vacuum state has the following properties: $a_S\left|0,0,0,0\right>$ $= a_I\left|0,0,0,0\right>$ 
$= a_R\left|0,0,0,0\right>$ $= a_D\left|0,0,0,0\right>=0$ (from which ``annihilation" operators) and $\left<0,0,0,0|0,0,0,0\right> = 1$ (inner product). Following the above definitions, we also have

\begin{eqnarray}
 &a^\dagger_S\left|n_S,n_I,n_R,n_D\right> =& 
 \left|n_S+1, n_I,n_R,n_D\right>~;\nonumber\\
 &a^\dagger_I\,\left|n_S,n_I,n_R,n_D\right> =& \left|n_S,n_I+1,n_R,n_D\right>~;\nonumber\\
 &a^\dagger_R\,\left|n_S,n_I,n_R,n_D\right> =& \left|n_S,n_I,n_R+1,n_D\right>~;\nonumber\\
 &a^\dagger_D\,\left|n_S,n_I,n_R,n_D\right> =& \left|n_S,n_I,n_R,n_D+1\right>~;\nonumber\\
&a_S\left|n_S,n_I,n_R,n_D\right> =& n_S\left|n_S-1, n_I,n_R,n_D\right>~;\nonumber\\ 
&a_I\left|n_S,n_I,n_R,n_D\right> =& n_I\left|n_S, n_I-1,n_R,n_D\right>~;\nonumber\\
&a_R\left|n_S,n_I,n_R,n_D\right> =& n_R\left|n_S, n_I,n_R-1,n_D\right>~;\nonumber\\
&a_D\left|n_S,n_I,n_R,n_D\right> =& n_D\left|n_S, n_I,n_R,n_D-1\right>~.
\label{eq5}
\end{eqnarray}

\vskip \baselineskip
\noindent
Since
\begin{eqnarray}
a^\dagger_S a_S\,\left|n_S,n_I,n_R,n_D\right> &=& n_S\left|n_S,n_I,n_R,n_D\right>~;\nonumber\\ 
a^\dagger_I a_I\,\left|n_S,n_I,n_R,n_D\right> &=& n_I\left|n_S,n_I,n_R,n_D\right>~;\nonumber\\
a^\dagger_R a_R\,\left|n_S,n_I,n_R,n_D\right> &=& n_R\left|n_S,n_I,n_R,n_D\right>~;\nonumber\\
a^\dagger_D a_D\,\left|n_S,n_I,n_R,n_D\right> &=& n_D\left|n_S,n_I,n_R,n_D\right>~,
\label{eq5b}
\end{eqnarray}

\vskip \baselineskip
\noindent
we may conclude that the operators $n_S=a^\dagger_S a_S$, $n_I=a^\dagger_I a_I$, $n_R=a^\dagger_R a_R$ and $n_D=a^\dagger_D a_D$ just count the number of individuals in a definite state, reason why they are called {\it number operators}.
Note that the vector state of our system may be then rewritten in terms of creation and annihilation operators as

\begin{eqnarray}
\left|\nu_S,\nu_I,\nu_R,\nu_D\right>_t &=&\sum_{n_S,n_I,n_R,n_D} p_{(n_S,n_I,n_R,n_D)}(t)\nonumber\\ &&\times\left(a^\dagger_S\right)^{n_S}\left(a^\dagger_I\right)^{n_I}\left(a^\dagger_R\right)^{n_R}\left(a^\dagger_D\right)^{n_D} 
\left|0,0,0,0\right>~. \label{eq6}
\end{eqnarray}

\vskip \baselineskip

\begin{figure}
    \centering
    \includegraphics[width=0.75\textwidth]{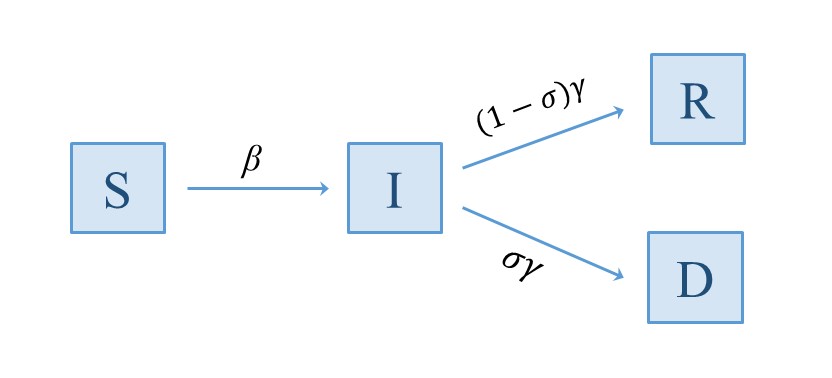}
    \caption{The structure of our SIR-type model. The rate $\beta$ is related to the infection rate of the virus and $\gamma$ is related to the time required to infected people to become recovered or deceased. Both parameters have the dimension of time$^{-1}$. Indeed, $\gamma=1/D$, where $D$ is the average time period during which an individual remains infectious, before one gets recovered or deceased. The parameter $\sigma$ (the case fatality rate) is dimensionless and represents the percentage of infected people who become deceased.}
    \label{SIRmodel}
\end{figure}

The time evolution of our system will then be generated by a linear operator $\mathcal{H}$ ({\it Hamiltonian}) which may be constructed directly from the transition rates present in our model according to Table \ref{table1} (cf. Ref.~\cite{MondainiBMST2015}, Table 1). In Figure \ref{SIRmodel} we present a compartmental diagram for the structure of our SIR-type model. 


The terms presented in Table \ref{table1} can be summed up and rearranged, so that the Hamiltonian can be written down as

\begin{eqnarray}
	\mathcal{H} = \beta n_S + \gamma n_I
       -\left[\beta a^\dagger_I a_S + (1-\sigma)\gamma a^\dagger_R a_I + \sigma\gamma a^\dagger_D a_I\right]~. \label{eq7}
\end{eqnarray}

\vskip \baselineskip
\noindent

\begin{table}
\caption{Transition rates presented in our model and corresponding terms in the Hamiltonian $\mathcal{H}$.}
\vskip \baselineskip
{\begin{tabular}{l l l}
\hline\hline
{Transition} & {Description} & {Contribution to $\mathcal{H}$}\\
\hline
$\mathcal{S} \xrightarrow{\beta} \mathcal{I}$ & infection (rate $\beta$)&$\beta(a^\dagger_S a_S-a^\dagger_I a_S)$\\
$\mathcal{I} \xrightarrow{(1-\sigma)\gamma} \mathcal{R}$ & change infectious $\rightarrow$ recovered (rate $(1-\sigma)\gamma$)&$(1-\sigma)\gamma(a^\dagger_I a_I-a^\dagger_R a_I)$\\
$\mathcal{I} \xrightarrow{\sigma\gamma} \mathcal{D}$ & change infectious $\rightarrow$ deceased (rate $\sigma\gamma$)&$\sigma\gamma(a^\dagger_I a_I-a^\dagger_D a_I)$\\
\hline
\end{tabular}}
\label{table1}
\end{table}

The notational advantage of this field theoretical description is made clear at this point if we observe that, from the above definitions, the equation which represents the flux of probability between states at rates defined by our model (the so-called {\it master equation} or 
{\it forward Kolmogorov differential equation} \cite{allen2010introduction}) takes the very compact form of an imaginary-time Schr\"odinger-type linear equation, namely

\begin{eqnarray}
\frac{d}{dt}\left|\nu_S,\nu_I,\nu_R,\nu_D\right>_t = -\mathcal{H}\left|\nu_S,\nu_I,\nu_R,\nu_D\right>_t~. \label{eq8}
\end{eqnarray}

\vskip \baselineskip
\noindent

We can get, after some algebra, a more common representation for the master equation by substituting the expressions for the Hamiltonian, 
(\ref{eq7}), and the vector state, (\ref{eq3}), into (\ref{eq8})

\begin{eqnarray}
\frac{d}{dt}\,p_{(n_S,n_I,n_R,n_D)}(t) &=& 
-(\beta n_s + \gamma n_I)\,p_{(n_S,n_I,n_R,n_D)}(t) \nonumber \\
&&+\beta(n_S+1)\,p_{(n_S+1,n_I-1,n_R,n_D)}(t)\nonumber\\
&&+ (1-\sigma)\gamma (n_I+1)\,p_{(n_S,n_I+1,n_R-1,n_D)}(t)\nonumber \\
&&+\sigma\gamma (n_I+1)\,p_{(n_S,n_I+1,n_R,n_D-1)}(t)\,~. \label{eq9}
\end{eqnarray}

\vskip \baselineskip

In order to compute the time-dependent expectation values of the observables $\mathcal{S}(t)$, $\mathcal{I}(t)$, $\mathcal{R}(t)$ and $\mathcal{D}(t)$ through the above master equation, 
we follow the well-established methodology presented in Ref.~\cite{allen2010introduction} and introduce the following {\it moment generating function (mgf)}

\begin{eqnarray}
M(\phi_S,\phi_I,\phi_R,\phi_D;t) &=& 
\left<e^{\phi_S \mathcal{S}(t)}e^{\phi_I \mathcal{I}(t)}e^{\phi_R \mathcal{R}(t)} e^{\phi_D \mathcal{D}(t)}\right>  \nonumber\\ &=&\sum_{n_S,n_I,n_R,n_D}\,p_{(n_S,n_I,n_R,n_D)}(t)e^{n_S\phi_S+n_I\phi_I+n_R\phi_R+n_D\phi_D}\,~.\label{eq10}
\end{eqnarray}

\vskip \baselineskip
\noindent
Note that from the above equation we have

\begin{eqnarray}
\left[\frac{\partial M}{\partial\phi_S}\right]_{\phi_S,\phi_I,\phi_R,\phi_D =0} &=& \sum_{n_S,n_I,n_R,n_D}n_S\,p_{(n_S,n_I,n_R,n_D)}(t) = \left<\mathcal{S}(t)\right>~;\nonumber \\ 
\left[\frac{\partial M}{\partial\phi_I}\right]_{\phi_S,\phi_I,\phi_R,\phi_D =0} &=& \sum_{n_S,n_I,n_R,n_D}n_I\,p_{(n_S,n_I,n_R,n_D)}(t) = \left<\mathcal{I}(t)\right>~; \nonumber \\ 
\left[\frac{\partial M}{\partial\phi_R}\right]_{\phi_S,\phi_I,\phi_R,\phi_D =0} &=& \sum_{n_S,n_I,n_R,n_D}n_R\,p_{(n_S,n_I,n_R,n_D)}(t) = \left<\mathcal{R}(t)\right>~; \nonumber \\ 
\left[\frac{\partial M}{\partial\phi_D}\right]_{\phi_S,\phi_I,\phi_R,\phi_D =0} &=& \sum_{n_S,n_I,n_R,n_D}n_D\,p_{(n_S,n_I,n_R,n_D)}(t) = \left<\mathcal{D}(t)\right>~.\label{eq11}
\end{eqnarray}

\vskip \baselineskip

Multiplying (\ref{eq9}) by $\exp(n_S\phi_S+n_I\phi_I+n_R\phi_R+n_D\phi_D)$ and summing on ($n_S,n_I,n_R,n_D$), leads, after some algebra, to

\begin{eqnarray}
\frac{\partial M}{\partial t} 
=  \left[\beta\left(e^{-\phi_S+\phi_I} - 1\right)\right]\frac{\partial M}{\partial\phi_S}
+ \left[\gamma\left(e^{-\phi_I+\phi_R} - 1\right)+\sigma\gamma\left(e^{-\phi_I+\phi_D} - e^{-\phi_I+\phi_R}\right)\right]\frac{\partial M}{\partial\phi_I}~.\label{eq13}
\end{eqnarray}

\vskip \baselineskip
\noindent
Hence, by  differentiating the above equation with respect to $\phi_S, \phi_I,\phi_R,\phi_D$ and evaluating the result at $\phi_S, \phi_I,\phi_R,\phi_D=0$ we get the following differential equations for $\left<\mathcal{S}(t)\right>$, $\left<\mathcal{I}(t)\right>$, $\left<\mathcal{R}(t)\right>$ and $\left<\mathcal{D}(t)\right>$:

\begin{eqnarray}
\left[\frac{\partial^2 M}{\partial t \,\partial \phi_S}\right]_{\phi_S, \phi_I,\phi_R,\phi_D=0} &=& \frac{d}{dt}\left<\mathcal{S}(t)\right> = -\beta \left<\mathcal{S}(t)\right>~;\nonumber \\
\left[\frac{\partial^2 M}{\partial t \,\partial \phi_I}\right]_{\phi_S, \phi_I,\phi_R,\phi_D=0} &=& \frac{d}{dt}\left<\mathcal{I}(t)\right> = \beta \left<\mathcal{S}(t)\right>-\gamma \left<\mathcal{I}(t)\right>~; \nonumber \\
\left[\frac{\partial^2 M}{\partial t \,\partial \phi_R}\right]_{\phi_S, \phi_I,\phi_R,\phi_D=0} &=& \frac{d}{dt}\left<\mathcal{R}(t)\right> = (1-\sigma)\gamma \left<\mathcal{I}(t)\right>~;\nonumber \\
\left[\frac{\partial^2 M}{\partial t \,\partial \phi_D}\right]_{\phi_S, \phi_I,\phi_R,\phi_D=0} &=& \frac{d}{dt}\left<\mathcal{D}(t)\right> = \sigma\gamma \left<\mathcal{I}(t)\right>~.
\label{eq14}
\end{eqnarray}

\vskip \baselineskip
\noindent
Last but not least, we must observe at this point that the above differential equations would be the same if we have considered time-dependent rates ($\beta(t)$,$\gamma(t)$) from the very beginning. This turns out to be an important aspect of our model as we shall see in the next section.

\section{Comparison to COVID-19 Epidemic Data - South Korea}
\label{modified}

We now compare our model against public epidemic data from South Korea, which was taken from the ``Worldometer" reference website \cite{worldometers}. The modeled data for the number of deaths and number of active cases was compared to the data reported in South Korea during 96 days of the epidemic. In order to take into account the slowdown of the spread due the effects of quarantine and social distancing we use a similar approach to Ref.~\cite{palladino2020modelling}, where the infection rate $\beta$ is a function of time ($\beta\equiv\beta(t)$). Specifically, $\beta(t)=\beta_0$ before a time $t_{th}$, while it exponentially decays for $t>t_{th}$. In our approach, however, we treat the period prior to lockdown and social actions with a a linearly growing $\beta(t)$ function, instead of a constant, i.e.:

\begin{eqnarray}
\beta(t)= \beta_0t  \ \ \
\mbox{for } 0 \leq  t < t_{th}~,
\end{eqnarray}

\vskip \baselineskip
\noindent
and due to function continuity the expression for $\beta(t)$ after lockdown measures becomes:

\begin{eqnarray} \label{eq:bettat}
\beta(t)= t_{th}\beta_0 \ e^{-(t-t_{th})/\tau}  \ \ \ \mbox{for } t \geq t_{th}~,
\end{eqnarray}

\vskip \baselineskip
\noindent
where the time $t_{th}$ represents the starting time of the quarantine actions, while $\tau$ refers to the decaying period and it has the dimension of time.


\section{Number of active cases}
\label{active}
For the number of active cases, we perform a weighted fit of the model allowing four free parameters to vary (the number of active cases is independent of $\sigma$, the case fatality rate, discussed in Section \ref{dead}) using least squares minimization with the LMFIT \cite{newville2016lmfit} package in Python 3.6.8. We use a population size of $N=5.164 \times 10^7$ people and as initial condition we use $19$ active cases, the number of registered cases on the 18th of February in South Korea according to the Worldometer database. The end date of this first wave of infections is set to May 23, 2020.

\begin{figure}[tb]
\centering
\includegraphics[width=0.75\textwidth]{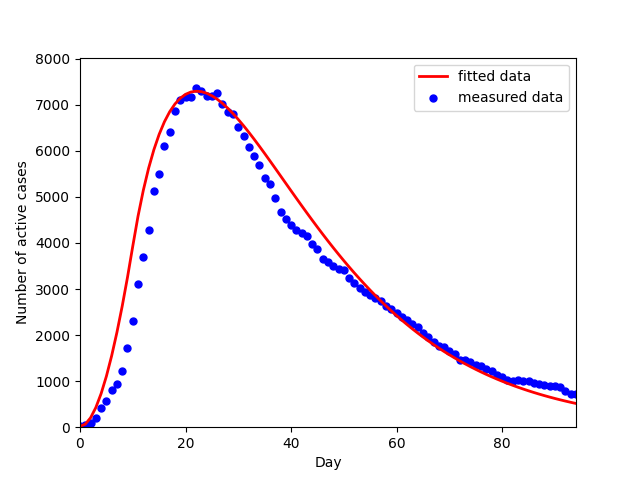} %
\caption{Number of active cases of COVID-19 in South Korea during 96 days of the epidemic corresponding to first wave of infections in the country. Modeled data is shown by the continuous red line, while measured data is depicted by the blue points.} 
\label{fig:active}
\end{figure}

The resulting fit, obtained by solving (\ref{eq14}) for $\left<\mathcal{I}(t)\right>$, overlaid with the real data points is shown in Figure \ref{fig:active}. We also obtain the mean absolute percentage error ($M$) of the real data and the fitted data points and achieve $M =  14.7\%$, which indicates acceptable agreement.

\subsection{Fitted parameters}

The best fit is given by the following set of parameters:

\begin{eqnarray}
 \beta_0 =1.898 \times 10^{-6} \pm 1.202 \times 10^{-7} \quad\quad &,& \quad\quad \gamma = 0.0710 \pm 0.0127~,\nonumber\\
 t_{th}= 10.008 \pm  0.868 \quad \quad&,& \quad \quad  \tau=19.010 \pm 3.029~,\nonumber
\end{eqnarray}

\vskip \baselineskip
\noindent
where the estimated uncertainties, calculated by LMFIT, are done by inverting the Hessian matrix, which represents the second derivative of fit quality for each variable parameter. It is interesting to note that 1/$\gamma$, the time required to infected people to become recovered or deceased, is estimated to be 1/0.0710 $\sim$ 14 days. The uncertainty on this value is, however, large (18\%), but even the shortest (12 days) and longest (17 days) values are within acceptable ranges for the average duration of the disease \cite{fernandez2020estimating}. 
The $\beta_0$ parameter, which is the initial constant for the infection rate, is a small number, which is unsurprising given that we have considered the infection rate to grow linearly with time until lockdown measures are in place. The value of $t_{th}$ is interpreted as the day lockdown and social distancing measures started in South Korea during the first wave of infections. 

\section{Case fatality rate}
\label{dead}

\begin{figure}[tb]
\centering
\includegraphics[width=0.75\textwidth]{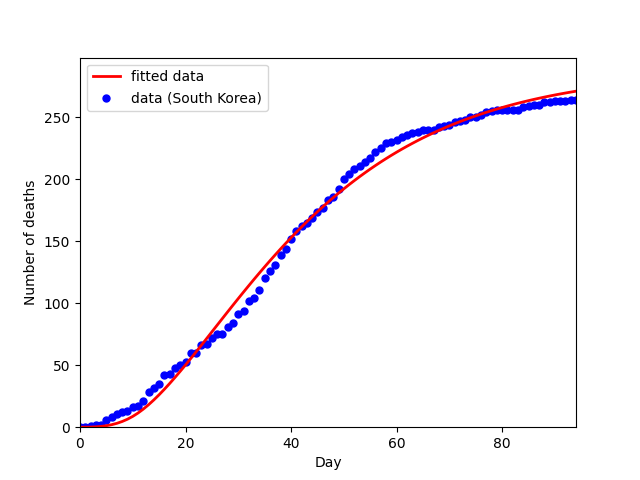} 
\caption{Number of deaths in South Korea during 96 days of the epidemic. Modeled data is shown by the continuous red line, while measured data is depicted by the blue points.} 
\label{fig:deaths}
\end{figure}

The case fatality rate ($\sigma$) in South Korea between February 18 and May 23, 2020 was obtained from the number of registered deaths. Figure \ref{fig:deaths} shows the modeled data (red), the solution for $\left<\mathcal{D}(t)\right>$ in (\ref{eq14}), overlaid with the measured data (blue), and reasonable agreement can be seen between them. The best fit of the measured data to our model is given by:

$$
  \sigma= 0.0109 \pm 0.0011~,
$$

\vskip \baselineskip
\noindent
which corresponds to a rate of $\sim 1.1\%$.

\section{Second and third outbreaks in South Korea}
\label{23outbreak}

\begin{figure}[t]
\centering
\includegraphics[width=0.75\textwidth]{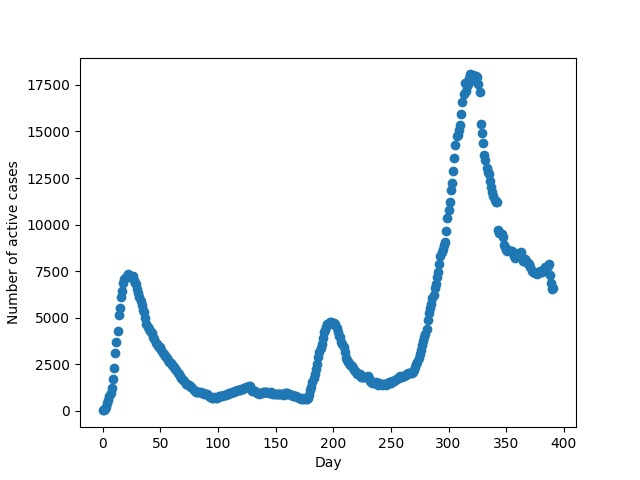} 
\caption{Number of active cases of COVID-19 in South Korea between February 18, 2020 and March 16, 2021. Three distinct waves of epidemic activity
can be clearly seen from the above data.} 
\label{fig:allwaves}
\end{figure}

\begin{figure}[t]
\centering
\includegraphics[width=0.75\textwidth]{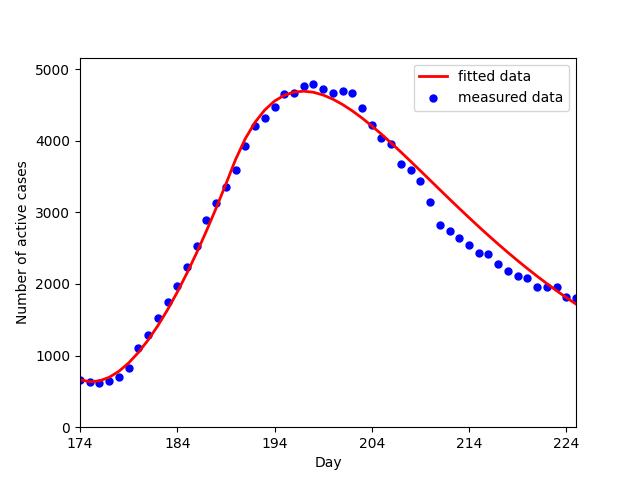} %
\caption{Number of active cases of COVID-19 in South Korea during 53 days of the epidemic corresponding to the so-called second wave of infections. Modeled data is shown by the continuous red line, while measured data is depicted by the blue points.} 
\label{fig:active2}
\end{figure}

South Korea has experienced three waves of the COVID-19 pandemic, as can be seen on Figure \ref{fig:allwaves}, which shows the total number of active cases as a function of time (days) using data taken from the Worldometer reference website. As our model was able to reasonably model the first wave of infections, it is worth checking its explanatory power during the second and third COVID-19 waves for the same country. There is no strict definition for what an epidemic wave is, with the term simply implying natural pattern of peaks and valleys of contaminated individuals. Thus, we define the second wave between August 10, 20120 
and October 1st 2020 
and fit the data to our model using the same approach as for the first wave. The result for the second wave can be seen on Figure \ref{fig:active2}, with the best fit given by the following set of parameters:

\begin{eqnarray}
 \beta_0=7.666 \times 10^{-7} \pm  2.086 \times 10^{-8} 
 \quad \quad &,& \quad \quad
 \gamma=  0.0741 \pm 4.110 \times 10^{-5}~, \nonumber\\
 t_{th}= 15.653 \pm  0.388
 \quad \quad &,& \quad \quad
 \tau=12.589 \pm 0.643~.\nonumber
\end{eqnarray}

\vskip \baselineskip


The mean absolute percentage error for the fit is of 5\%, which points to an even better agreement compared to the first wave, a fact that is partially explained by the better available statistics. It must also be noted that the end of the second wave of infections quite rapidly became the start of the third wave, reason the data was truncated already on October 1st, i.e., equation \ref{eq:bettat} no longer holds, after around that date. The good agreement during the expanding phase, followed by a worse agreement during the decay phase seen in Figure
\ref{fig:active2} can be partially attributed to it.

\begin{figure}[tb]
\centering
 \includegraphics[width=0.75\textwidth]{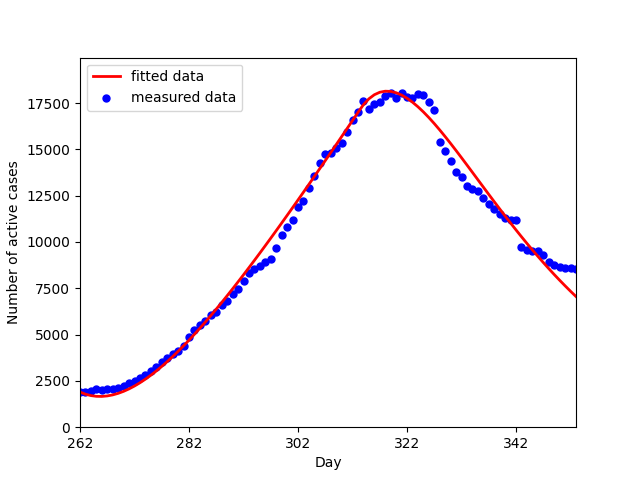} %
\caption{Number of active cases of COVID-19 in South Korea during 93 days of the epidemic corresponding to the so-called third wave of infections. Modeled data is shown by the continuous red line, while measured data is depicted by the blue points.} 
\label{fig:active3}
\end{figure}

Figure \ref{fig:active3} shows the result for the third epidemic wave in South Korea, between November 06 2020 and February 06 2021. The $\beta_0$ parameter is of the same order of magnitude as for the second wave and there is also similar agreement, with a mean absolute percentage error of 5.7\%.

\section{Discussion of the model approach and results}
\label{discussion}
The second quantization approach described in the present article is generic and can be applied  to a variety of problems for which stochastic modeling is appropriate, being the main constraint of its application to classical systems, as opposed to quantum systems. In, e.g., Ref.~\cite{MondainiBMST2015} this approach was applied to the case of hepatitis C virus epidemic, but no confrontation with real data was possible, and as such, it served only as an initial presentation of the theoretical framework and a glimpse of its full potential. To further reinforce the generality of the second quantization approach to stochastic models it is also worth mentioning Ref.~\cite{Mondaini_2017}, which presented a simple two-variable (healthy/cancer cells) system to the description of tumour growth, but again without yet successful confrontation with real data. The present article presents therefore, not only an application of the theoretical framework established in Ref.~\cite{MondainiBMST2015} to a new problem, but also its first successful description of real data.

Moreover, on the specific problem studied in the current article, namely the COVID-19 epidemic, the model was successfully expanded by the use of a time-dependent rate of infection, with a linear growing function for the initial phase of the infection rate representing the most appropriate way to describe the data. This is interesting from a public policy perspective, since it reinforces the social distancing measures in the control of the COVID-19 epidemic. According to our model, on the absence of controls, the rate of infection itself is accelerating.

Finally, the tension observed between model and data during the second and third waves of the epidemic can, to a reasonable degree, be attributed to the fact that the rate of infection decrease was not truly exponential, as observed in the first wave. Again this might also have some interest for policy makers since a quantifiable disagreement with an exponential decrease, when using our model, indicates a potential change in the mathematical behavior of the infection rate, which can quickly shift the epidemic phase from infection decline to infection grow.

\section{Concluding Remarks}
\label{conclusion}

Using an SIR-type stochastic model built on field theory techniques, we have investigated the  time evolution of the mean number of infectious (active cases) and deceased individuals for COVID-19 epidemic in South Korea. Our model was able to describe the available data successfully. To the best of our knowledge, it is the first COVID-19 model built from standard field theoretical language. 
Despite its simplicity, the model was able to describe three waves of infections of the COVID-19 pandemic in South Korea, which indicates its robustness and versatility. 
Studies including a non-susceptible subset of the population due to vaccinations and their effects on the infection rates, as well as comparisons with data from other countries, are possible future directions to be explored with our model.



\vskip \baselineskip
\vskip \baselineskip

\References
\bibliographystyle{iopart-num}

\item
\noindent
\bibliography{covid.bib}

\endrefs
\end{document}